\documentclass{emulateapj}

\shorttitle{AEOS Adaptive Optics PSF}
\shortauthors{Makidon et al.}

\newcommand \conv   {\mathbf{\ast}}

\newcommand \eg     {{\it e.g., }}

\newcommand \ie 	{{\it i.e., }}

\newcommand \eq     {\,=\,}                 

\newcommand \Iband  {$I$-band}
\newcommand \Hband  {$H$-band}
\newcommand \Kband  {$K$-band}
\newcommand \hne    {h_{NE}}
\newcommand \hnw    {h_{NW}}
\newcommand \hse    {h_{SE}}
\newcommand \hsw    {h_{SW}}
\newcommand \Shah	{\rm III}

\begin{document}

\title{An Analysis of Fundamental Waffle Mode in Early AEOS Adaptive Optics
	Images\footnote{Based on observations made at the Maui Space Surveillance
	System operated by Detachment 15 of the U.S. Air Force Research
	Laboratory's Directed Energy Directorate.}}

\author{Russell B.\ Makidon\altaffilmark{1}, Anand Sivaramakrishnan\altaffilmark{1}}
\affil{Space Telescope Science Institute, 3700 San Martin Drive,
	Baltimore, MD 21218}
\email{makidon@stsci.edu}

\author{Marshall D.\ Perrin\altaffilmark{1,2}}
\affil{Astronomy Department, 601 Campbell Hall, University of California,
	Berkeley, CA 94720}
	
\author{Lewis C.\ Roberts, Jr}
\affil{The Boeing Company, 535 Lipoa Parkway, Suite 200, Kihei, HI 96753}

\author{Ben R.\ Oppenheimer}
\affil{Department of Astrophysics, The American Museum of Natural History,\\
	Central Park West at 79th Street, New York, NY 10024}

\author{R\'{e}mi Soummer\altaffilmark{1,3}}
\affil{Space Telescope Science Institute, 3700 San Martin Drive,
	Baltimore, MD 21218}

\and

\author{James R.\ Graham\altaffilmark{1}}
\affil{Astronomy Department, 601 Campbell Hall, University of California,
	Berkeley, CA 94720}

\altaffiltext{1}{NSF Center for Adaptive Optics}
\altaffiltext{2}{Michelson Graduate Fellow}
\altaffiltext{3}{Michelson Postdoctoral Fellow}

\begin{abstract}

Adaptive optics (AO) systems have significantly improved astronomical
imaging capabilities  over the last decade, and are revolutionizing the
kinds of science possible with 4-5m class ground-based telescopes.  
A thorough understanding of AO system performance at the telescope can
enable new frontiers of science as observations push AO systems to their
performance limits.
We look at recent advances with wave front reconstruction (WFR) on
the Advanced Electro-Optical System (AEOS) 3.6 m telescope to show how
progress made in improving WFR can be measured directly in improved
science images.  We describe how a ``waffle mode'' wave front error
(which is not sensed by a Fried geometry Shack-Hartmann wave front sensor)
affects the AO point-spread function (PSF). We model details of AEOS AO
to simulate a PSF which matches the actual AO PSF in the \Iband, and show
that while the older observed AEOS PSF contained several times more waffle
error than expected, improved WFR techniques noticeably improve AEOS AO
performance.   We estimate the impact of these improved WFRs on \Hband\
imaging at AEOS, chosen based on the optimization of the Lyot Project
near-infrared coronagraph at this bandpass.

\end{abstract}

\keywords{instrumentation: adaptive optics -- 
          instrumentation: miscellaneous --
          techniques: miscellaneous -- 
          methods: numerical }

\slugcomment{Accepted by PASP}

\section{Introduction} \label{sec:Introduction}

Various wave front sensors (WFS's) are in principle insensitive to certain
kinds of wave front errors.  The commonly-used square `Fried geometry' 
sensing \citep{FriedGeom77} in most Shack-Hartmann-based adaptive optics 
(AO) systems is insensitive to a checkerboard-like pattern of phase error 
called waffle.  This zero mean slope phase error is low over one set of 
WFS subapertures arranged as the black squares of a checkerboard, and 
high over the white squares.  

Since waffle modes are in the null space of a Fried geometry WFS, they
can be present in the reconstructed wave front unless the wave front
reconstructor (WFR) explicitly removes or attenuates these modes.  
Because waffle is so common to Shack-Hartmann systems --- early images
from the 241-actuator Palomar AO system \citep{Troy00}, the 349-actuator
Keck AO system \citep{Wizinowich2000b}, and the 941-actuator Advanced
Electro Optical System (AEOS) AO system \citep{Roberts02} all showed the
presence of significant waffle --- effort has been made to understand 
how to control this and other ``blind'' reconstructor modes
\citep{GavelSPIE03,PoyneerSPIE03,MakidonSPIE03}.  Advances in wave front
reconstruction approaches have improved AO system performance considerably
on all three of the aforementioned telescopes (M.~Troy 2003, {\it private
communication}; D.~Gavel 2003, {\it private communication}; G.~Smith 2003, 
{\it private communication}).

However, the prevalence of waffle in most Shack-Hartmann-based AO
systems led us to investigate the physical nature of waffle, without a
sophisticated control-theoretic formalism.  Here we present the
understanding we gained about waffle mode error using a combination of
Fourier analysis and numerical simulation to examine the effects of waffle
on the observed point spread function (PSF).  We concentrate our efforts
on the fundamental waffle mode, where the checkerboard extends over the
full aperture of the telescope (a complete enumeration of all waffle modes
can be found in \citet{GavelSPIE03}). 

We follow the methods of \cite{Sivaramakrishnan01} (hereafter referred to
as SKMBK), which established simulations calibrated to AO imaging data
from Palomar as a viable means to examine AO system characteristics.  We
have tuned our simulations to the AEOS Adaptive Optics PSF in an attempt
to reproduce waffle behavior seen in early data acquired with this system. 

The 3.6 m AEOS telescope, part of the U.S. Air Force Research Laboratory's
Maui Space Surveillance System (MSSS), is arguably the best telescope on 
which to
develop the nascent field of Extreme Adaptive Optics (ExAO).  This is a
result both of the telescope's location at a prime astronomical site
possessing good seeing and, more importantly, the AEOS adaptive optics
system's 941-actuator deformable
mirror, with 35 actuators spanning the diameter of the DM \citep{Roberts02}.
This provides the highest actuator density available to any civilian
astronomical observing program, with a projected actuator spacing of
order 0.11 m/actuator at the primary mirror.

In Section~\ref{sec:WaffleError} we examine the nature of waffle mode error
using Fourier analysis and the PSF expansion of \citet{Sivaramakrishnan02,
Perrin03}.  We apply this understanding to simulations of waffle mode in 
AEOS \Iband\ images in Section~\ref{sec:simulations}.  We then extrapolate 
our \Iband\ imaging simulations to \Hband\ in Section~\ref{sec:NIR} to
predict AO system performance in the near-infrared at AEOS.  Our choice in
simulating \Hband\ is predicated on the Lyot Project
coronagraphic imager's optimization for operations at \Hband\ and the 
scientific potential of this instrument \citep{OppenheimerSPIE04}.\footnote{See also: http://www.lyot.org}

\section{The Nature of Waffle Mode Error}  \label{sec:WaffleError}  

Waffle mode error is inherent in any wave front sensing that uses a 
square array of wave front slope sensors. The atmospheric phase 
disturbance typically possesses some power in this mode, which will 
``leak'' through the AO system to produce a square grid of ghost images 
centered around the true PSF peak. We examine this process in detail,
first for a simplified case consisting of a single WFS cell and then 
extending to the full sensor array.

The Shack-Hartmann WFS geometry is the most common choice for AO system
with more than a few tens of actuators.  Its choice is driven by the
essentially square nature of WFS detector pixel layout, as well as the 
relative simplicity of the concept.  Another type of WFS, the curvature
sensor \citep{Roddier88}, is not sensitive to waffle mode error,
although it has its own particular null space.  Curvature sensors are 
typically not used for AO systems with more sensing channels because of 
their noise propagation properties \citep{Glindemann00}. 
AO systems at Keck \citep{Wizinowich2000a, Wizinowich2000b}, Palomar 
Hale \citep{Troy00}, and AEOS \citep{Roberts02} all use 
Shack-Hartmann WFSs, imaging light onto a CCD detector using square
lenslet arrays located at an image of the entrance pupil (the Shane 3~m
telescope at Lick Observatory is an exception, having a hexagonal grid
Shack-Hartmann WFS geometry). The WFS detector
is read out, and the on-chip $x$ and $y$ coordinates of the spots'
centroids are delivered to the wave front reconstruction computer. The
difference of one of the centroid's coordinates from its long-term
temporal mean value is directly proportional to the slope ${\bf s} =
(s_x, s_y)$ of the wave front over the corresponding subaperture. The
constant of proportionality is just the plate scale of the WFS
lenslet.  We can therefore use the quantity 
${\bf s} = (s_x,s_y) = (\kappa(x_c - \langle x_c\rangle), 
	\kappa(y_c - \langle y_c\rangle))$
as the measured quantity in any given subaperture.  Here $(x_c,y_c)$ is
the location of an instantaneous centroid of a spot in the
Shack-Hartmann subaperture (in pixels), $\langle\rangle$ denotes
temporal averaging, and $\kappa$ is the plate scale of the detector (in
arcseconds or microradians per pixel).  This gives us the vector slope
of the wave front (in angular units) on a grid of locations over the
full aperture.

\subsection{The unit cell} \label{sec:cell}

\begin{figure}
	\epsscale{0.55}
	\plotone{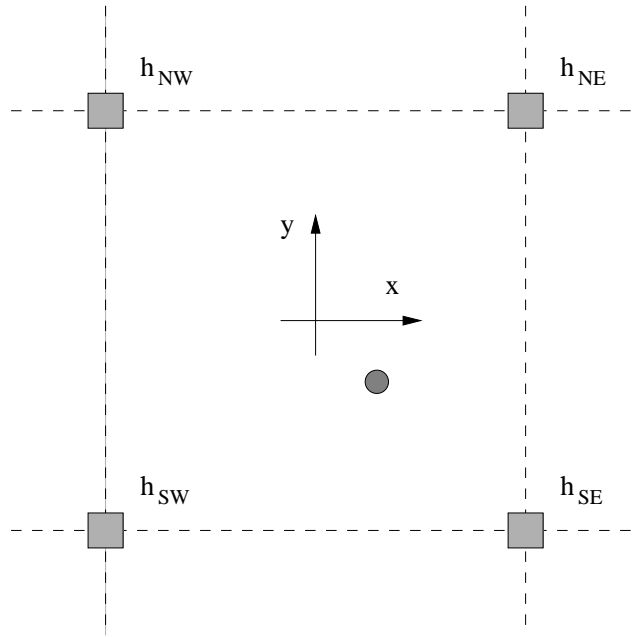}
	\caption{
	  The unit cell of the Fried geometry wave front sensor. Actuators
	  are shown as solid squares, and lenslet edges as dashed lines. 
	  The $(x,y)$ coordinate system is in wave front sensor pixels.  The
	  centroid of the image produced by the lenslet is shown as the
	  dark circle.  The subaperture, when projected back on to the
	  primary, is a square of side $a$.
	}
	\label{fig:cell}
\end{figure}

The nature of the smallest spatial scale waffle mode can be deduced 
from inspecting a `unit cell' of the WFS array.  We assume a WFS 
subaperture which has four DM actuators, one at each corner of a square 
(see Figure~\ref{fig:cell}).  The relationship between the wave front slope 
across this subaperture and the heights of wave front at the four corners 
of the square contains the key to the waffle mode error, since the DM
actuators at these corners will need to be moved by exactly half these 
heights to flatten the wave front across the subaperture.  Of course 
the actuators could be placed anywhere relative to the subaperture grid, 
though real AO systems typically try to maintain the DM actuator-to-pupil 
alignment we describe if the actuator spacing and geometry matches the 
subaperture geometry.  We stress here that {\it waffle mode error does 
not depend fundamentally on the actuator location and geometry:} it 
only depends on the sensor geometry.

We label the corners of the subaperture with mnemonics generated from 
the cardinal directions, assuming North points toward the top of the
figure.  The corner heights are therefore
	$ {\bf h} = (\hne, \hnw, \hse, \hsw)$.
We assume that the wave front is flat across the
subaperture, so we can immediately deduce the way our
slope measurement ${\bf s}$ constrains the wave front heights
at the corners,
\begin{eqnarray*}
        (\hne + \hse)/2d - (\hnw + \hsw)/2d &=&  s_x  \\
        (\hnw + \hne)/2d - (\hse + \hsw)/2d &=&  s_y,
\end{eqnarray*}
where $d$ is the subaperture spacing as well as the inter-actuator spacing.
These are the two equations of condition of the problem.
This relation between the four unknowns {$\bf h$} and the two
measured quantities {$\bf s$} can be re-expressed as
\begin{eqnarray*}
        \hse - \hnw  &=&  d(s_x - s_y)  \\
        \hne - \hsw  &=&  d(s_x + s_y).
\end{eqnarray*}
We use our only two observable quantities, $s_x$ and $s_y$, to relate
the heights across the diagonals of the subaperture. This leaves two
undetermined quantities in these equations relating our unknown
heights to the two data points: the equations of condition are
rank-deficient by~2.

It is not possible to determine all four heights in $\bf h$ with just
two observables.  If we face the engineering problem of having to
command four actuators based on the data contained in $\bf s$, we will
have to add two arbitrary equations to the defining equations of the
problem to solve them. That is, we have to add two more equations
to the system of equations in order to obtain a solution for ${\bf h}$.  
One such equation  is easy to come up with: the physics of the problem 
indicates that the mean value of the four heights can be any value we 
choose, as it is the piston of the wave front.  We can therefore add 
the constraint equation
\begin{equation} \label{piston}
          (\hne + \hnw + \hse + \hsw)/4 \eq  P
\end{equation}
to the equations of condition.  Here $P$, the mean height of the wave front,
is arbitrary. 

Our original defining equations are difference equations, so it is scarcely 
surprising that the offset of the heights is unknown.  Wave front slope 
data alone cannot determine the piston of the wave front.  Let us assume 
a zero piston in our case (we will set the mean voltages sent to the DM 
actuators using common sense in any real system):
\begin{equation} \label{zeropiston}
          \hne + \hnw + \hse + \hsw \eq  0.
\end{equation}

We still need one more equation without data --- a constraint
equation --- to solve our problem.  This is the equation that determines
how much waffle we introduce into our very simple AO system.  So far we
have set the slopes across the unit cell diagonals using data, and the
mean piston	using common-sense.  We must now guess how the means of the
two sets of	diagonally-related actuators differ from each other, in the
absence of new information (data).  If the wave front is relatively flat
across the subaperture (\eg~ $d \ll r_0$, where $r_0$ is the Fried
length \citep{FriedLength66}), we can assert that the means
are equal.  This statement,	expressed as the equation
\begin{equation} \label{waffleconstraint}
          (\hne + \hsw) -  (\hnw + \hse) \eq  0,
\end{equation}
adds the equation required to solve the system of
equations relating our two WFS slopes to our four actuator heights 
completely. It determines how much waffle mode present in the atmosphere
we let slip through our AO system, on average. Equation~(\ref{waffleconstraint})
is an {\it equation of constraint}, since it does not involve any data.
We note that ellipticity of the individual WFS subapertures's spots can 
yield data on the waffle mode error, although no AO system we know of uses
this information in practice.

A singular value decomposition (SVD) method is often used to solve the
underdetermined set of equations between WFS data and actuator heights.
The physics of what really happens when we perform a wave front 
reconstruction with SVD is not apparent:  this
method does not distinguish between constraints and equations of condition.

\subsection{The full lattice} \label{sec:lattice}

Here we extend the description in Section \ref{sec:cell} 
to the simplest case when we have more data than unknowns.  We will 
use a square aperture. While this is obviously unrealistic, our 
analysis of this case can be applied to arbitrary aperture geometries, 
and leads to an understanding of the cause of unsensed modes in wave 
front reconstructors that use slope measurements as input to their 
algorithms.

\begin{figure}
	\epsscale{0.75}
	\plotone{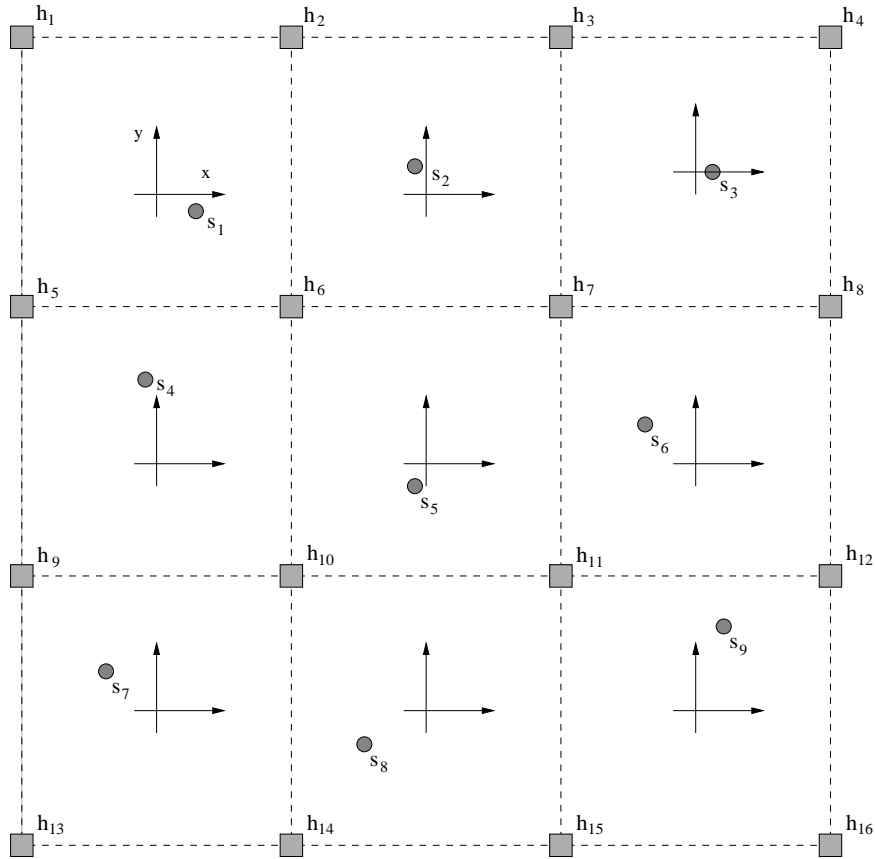}
	\caption{
	  A square aperture Fried geometry wave front sensor array, with
	  sixteen actuators (shown as squares, each at a height $h_{i}$)
	  and nine wave front sensors.	Wave front slopes $s_{i} =
	  (s_{xi},s_{yi})$ denote the instantaneous centroids of a lenslet
	  spot in the Shack-Hartmann subaperture (in on-chip pixels).
	}
	\label{fig:array}
\end{figure}

Any Fried geometry WFS system on a simply-connected aperture will
possess two families of actuator heights.  Extending the unit cell to a
case when there are more data points than unknown actuator heights
demonstrates this.  The $N_s$ WFS subapertures produce $2N_s$ slope
measurements, since each subaperture yields an $x$ and a $y$ slope. 
With a square aperture, the simplest such case is that of a grid of
four actuators on a side, enclosing a square of subapertures three to a
side  (see Figure~\ref{fig:array}). We must deduce 16 actuator heights
using 18 pieces of slope data. On the face of it the problem appears to
be overdetermined, but is in fact still underdetermined. Writing out the
equations that relate slopes to actuator heights explicitly, the set of
equations governing one family of heights is
\begin{eqnarray}
	h_6     &=&  	h_1    - \delta_1   \nonumber \\
	h_3     &=& 	h_6    + \sigma_2   \nonumber \\
	h_9     &=& 	h_6    - \sigma_4   \nonumber \\
	h_8     &=&  	h_3    - \delta_3   \nonumber \\
	h_8     &=&  	h_{11} + \sigma_6   \nonumber \\
	h_{11}  &=& 	h_6    - \delta_5   \nonumber \\
	h_{11}  &=& 	h_{14} + \sigma_8   \nonumber \\
	h_{14}  &=& 	h_9    - \delta_7   \nonumber \\
	h_{16}  &=& 	h_{11} - \delta_9,  
\end{eqnarray}
and the equations for the second family are
\begin{eqnarray}
	h_2     &=& 	h_7    + \delta_2   \nonumber \\
	h_5     &=& 	h_2    - \sigma_1   \nonumber \\
	h_7     &=& 	h_4    + \sigma_3   \nonumber \\
	h_{10}  &=& 	h_5    - \delta_4   \nonumber \\
	h_{10}  &=& 	h_7    - \sigma_5   \nonumber \\
	h_{12}  &=& 	h_7    - \delta_6   \nonumber \\
	h_{13}  &=&  	h_{10} - \sigma_7   \nonumber \\
	h_{15}  &=& 	h_{12} - \sigma_9   \nonumber \\
	h_{15}  &=& 	h_{10} - \delta_8, 
\end{eqnarray}
where
\begin{eqnarray} 
	\delta_i & = & d({s_i}_y  - {s_i}_x) \nonumber \\
	\sigma_i & = & d({s_i}_y  + {s_i}_x).
\end{eqnarray}

These two sets of nine equations each are disjoint --- no individual
unknown	actuator height appears in both sets.  We see by inspection 
that we can	choose any one of the first set of heights arbitrarily.
Setting $h_1$ to a fixed number sets every actuator height in the first
set. Similarly, we can choose $h_2$ to be any value we please. Each of
the two families of actuator heights therefore has an arbitrary net
piston of its own (\ie~ the mean of	all the actuator heights in the
family).  This is the common-sense expression of the statement that the
matrix describing the system is	rank-deficient by 2. These
pistons can individually be set to be equal, thereby selecting a
`waffle-free' solution.  This solution still has arbitrary total piston.
Again, we can select a zero piston solution from all the waffle-free
solutions by requiring the pistons of each family be zero:
\begin{equation}
	\sum \limits_{p=1}^{16} h_p  = 0.
\end{equation}
The atmospheric phase disturbance at any given instant does not
typically possess zero waffle when viewed through the WFS subapertures.
That is, if one regards the square WFS lenslets (projected back to the
entrance pupil) as a checkerboard, the mean phase over the black squares
will not equal the mean phase over the white squares. However, as
\citet{GavelSPIE03} shows, for subapertures that are small compared
to $r_0$ the net waffle will be small compared to unity (waffle mode
error has units of radians, since it is	a mean of a phase).	By enforcing
a zero-waffle solution on our actuator heights we introduce some error
in our fit of the wave front. This error produces the familiar waffle
pattern appearing at an angular radius of $\sqrt 2 (\lambda / 2d)$ in
the image (see Section~\ref{sec:simulations}).

\subsection{Waffle Mode Error and the PSF} \label{sec:wafflePSF}

Here we consider waffle mode error as a phase aberration
$\phi_{w}(x,y)$ over the pupil, and trace the effect of the waffle mode
on the AO-corrected monochromatic wave front analytically.  We use the
notation of \citet{Sivaramakrishnan02}, which we restate here.

The telescope entrance aperture and all phase effects in a
monochromatic wave front impinging on an optical system can be
described by a real aperture illumination function $A(x,y)$ multiplied
by a unit modulus function $A_{\phi}(x,y) = e^{i\phi(x,y)}$.  The
aperture plane coordinates $(x,y)$ are in units of the wavelength of the
monochromatic light under consideration (to simplify the Fourier
transforms), while the corresponding image
plane coordinates $(\xi,\eta)$ are
in radians. Here, phase variations induced by the atmosphere or
imperfect optics  are described by a real wave front phase function
$\phi$ which possesses a zero mean value over the aperture plane, 
\begin{equation} \label{eqn:PhiMeanZero}
	\frac{\int A\,\phi\,dx\,dy} {\int A\,dx\,dy}= 0.
\end{equation}
We chose the location of the image plane origin at the centroid of the
image PSF, which corresponds to a zero mean tilt of the wave front over
the aperture \citep{Teague82,Sivaramakrishnan95}.

We assume that the transverse electric field in the image plane is
given by the Fourier transform (FT) of the field in the aperture plane
\citep{Goodman95}.  We indicate transform pairs by a change of case,
where the FT of $A$ is $a$ and the FT of $\phi$ is $\Phi$.

The ``amplitude spread function'' (ASF) of an optical system with phase
aberrations is the FT of $A_{\psi} = AA_{\phi}$, and is given by
$a_{\psi} = a\ \conv\ a_{\phi}$, where $\conv$ denotes convolution.
This complex-valued quantity is proportional to the amplitude of the 
incoming wave in the image plane.  The complex number's phase at any point
is interpreted as a relative phase difference in the wave front at that point
in the image plane.
The corresponding PSF of this optical system is 
$p_{\psi}^{\ } = a_{\psi}^{\ } a_{\psi}^{\ast}$,
where the superscripted $ ^{\ast}$
indicates complex conjugation (see \eg \citet{Bracewell00} for the
fundamental Fourier theory and Fourier optics conventions).

For this analysis, we assume that the total phase $\phi(x,y)$ is given by
$\phi_{AO}(x,y) + \phi_{w}(x,y)$, where $\phi_{AO}(x,y)$ is the residual
atmospheric phase after correction by the AO system.  Thus the unit modulus 
function $A_{\phi}(x,y)$ can be written as
\begin{equation} \label{eqn:UnitModulus}
	A_{\phi} = A_{AO}A_{\phi_{w}} = 
	    e^{i\left [\phi_{AO}(x,y)+\phi_{w}(x,y)\right ]}.
\end{equation}
In order to understand the effects of waffle mode error on the PSF,
we now look at a waffle aberration without any atmospheric aberration
contribution. Setting $\phi_{AO}\eq 0$, the aperture illumination function reduces to
\begin{equation} \label{eqn:WaffleApertureIllum}
	A_{\psi} = A\,A_{\phi_{w}} = A\,e^{i\phi_{w}(x,y)}.
\end{equation}

In Section~\ref{sec:lattice}, we presented waffle mode error as a
checkerboard pattern in phase with a period of $2d$, where $d$ is the
WFS subaperture size in the Fried geometry. In the one-dimensional case,
we write this as a top-hat function $h\Pi(x/d)$ convolved with
series of Dirac $\delta$-functions spaced at a period $2d$.
Here $h$ is a function of the atmospheric phase $\phi_{atm}$, defining
the ``strength" of the waffle function.  The periodic function describing
our checkerboard thus takes the form
\begin{equation} \label{eqn:WaffleFunction}
	{\Shah}\left ( \frac{x}{2d} \right ) \conv\ 
		h\Pi\left (\frac{x}{d}\right ) = 
		2dh\sum_{n} \Pi\left (\frac{x}{d} - 2nd\right ),
\end{equation}
where we use the shah function, ${\Shah}(x/2d)$ to denote a series
of Dirac $\delta$-functions at the required spacing \citep{Bracewell00}.
The waffle phase error $\phi_{w}$ can then be written
\begin{equation} \label{eqn:phiWaffle}
	\phi_{w} = \Pi\left (\frac{x}{D}\right ) \cdot 
		\left [{\Shah}\left ( \frac{x}{2d} \right ) \conv\ 
		h\Pi\left (\frac{x}{d}\right )\right ],
\end{equation}
where $D$ is the telescope aperture diameter in units of wavelength. 
The wave amplitude in the image plane due to waffle mode is the FT 
of $\phi_{w}$:
\begin{equation} \label{eqn:FTphiWaffle}
	\Phi_{w} = hD\ {\rm sinc}(D\theta)\ \conv\ 
		\left[2d\ {\Shah\,(2d\theta)} \cdot d\ {\rm sinc}(d\theta)\right].
\end{equation}
The above expression is the field due to the telescope aperture convolved 
with a sinc function sampled at $\theta = n/2d$, where $n$ is an integer.
However, the function 
${\rm sinc}(d\theta)$ is zero for each even value of $n$ (\eg $n=2,4,6,\dots$).
Thus this fundamental waffle mode error results in the replication of 
the perfect PSF at a lattice of points at the angular period equal to the
spatial Nyquist frequency of the WFS subaperture density, the relative
intensity of each point being modulated by ${\rm sinc}(d\theta)$ (with half
of those points falling on the zeros of the ${\rm sinc}(d\theta)$).  

\subsubsection{The PSF Expansion} \label{sec:WafflePSFexpansion}

\citet{Perrin03} developed a general form for an expansion of an aberrated
PSF, $p$, in terms of the Fourier transform of the phase aberration 
$\phi$ over the aperture.  They express the PSF was given as a sum of 
individual terms in an infinite convergent series, with the $n^{\rm th}$ term of 
that series given by
\begin{equation} \label{eqn:PSFExpansion}
	p_{n} = i^{n}\sum_{k=0}^{n} \frac{(-1)^{n-k}}{k!(n-k)!}
		(a\ \conv^{k}\ \Phi)(a^{\ast} \conv^{n-k}\ \Phi^{\ast}),
\end{equation}
where $\conv^{n}$ is used to denote an $n$-fold convolution operator.
We refer the reader to the discussion of the second-order expansion of
the partially corrected PSF in \citet{Sivaramakrishnan02} and in 
\citet{Perrin03}, noting here that while the first order term
\begin{equation} \label{eqn:PSF_n1}
	p_{1}^{\ } = i\Bigl[a^{\ast}(a\ \conv\ \Phi) -
		a(a^{\ast}\ \conv\ \Phi^{\ast})\Bigr]
\end{equation}
is due entirely to the antisymmetric component of $A_{\psi}$, it does not
contribute significantly to the waffle pattern.  This is because the 
function $a(x,y)$ is very small at points beyond a few $\lambda/D$ from 
the center of the on-axis PSF, and the first order term is multiplied by 
$a$ (see \citet{Bloemhof01,Sivaramakrishnan02} for further discussion
of pinned speckles).  This is borne out by simulations with pure 
waffle-mode phase aberrations showing the symmetric terms of the PSF 
expansion (\eg the $p_{2}$ and $p_{4}$ terms of \citet{Perrin03}) 
dominate, even when the 
phase error itself is large and antisymmetric. We discuss this further 
in Section \ref{sec:WaffleStrehl}.

\section{Simulations} \label{sec:simulations}

Our goal in this analysis is to understand the current performance of the 
AEOS AO system in the \Iband, and to enable predictions of AO system
performance in 
the \Hband. Such predictions are useful for determining the contrast
ratios accessible to the Lyot Project AO-optimized near-infrared (NIR)
coronagraph on AEOS. Our simulations incorporate the AEOS pupil geometry, 
namely a 3.63 m clear aperture for the primary mirror, with a 0.40~m
obstruction from the secondary and tertiary mirrors. We do not 
model secondary support structures.

A detailed simulation of an AO system including all aspects of
wave front sensing, reconstruction, and correction requires substantial
computational effort.  For some purposes it is sufficient to model the
action of an AO system as a transfer function or spatial frequency filter
acting on the incident wave front.  In SKMBK, we used a 
high pass filter in spatial frequency space for this purpose.
We tuned our initial model to fit data from the Palomar AO system reported in
\citet{OppenheimerSPIE00}.  We briefly describe the SKMBK model here, 
repeating some of the arguments presented in that work, and show how we 
modified the model to fit the AO imaging data from AEOS.

The cutoff spatial frequency of the AO filter cannot be higher than
the spatial Nyquist frequency of the actuator spacing, which is
\begin{equation} \label{k_AO}
    k_{AO} = N_{act}/2D = 1/2d,
\end{equation}
where $N_{act} = D/d$, with $D$ the primary mirror diameter and $d$ the
actuator spacing projected onto the primary mirror. 
At a wavelength $\lambda$, this spacing in the pupil plane
corresponds to a limiting angle of correction in the image plane,
\begin{equation} \label{thetaAO}
    \theta_{AO} = N_{act} \lambda / 2 D,
\end{equation}
which defines the `AO control radius' (\citet{Perrin03,Poyneer04} discuss
the AO control radius in more detail).

As we note in SKMBK, the shape of the AO filter near the cutoff 
frequency depends on details of the AO system. In general, DM actuator
influence functions which extend to neighboring actuators' positions 
reduce the sharpness of the cutoff. Noisy wave front sensing and 
intrinsic photon noise reduce the efficacy of high spatial frequency wave
front correction. The flow of the atmosphere past the telescope pupil 
and a non-isotropic refractive index spatio-temporal distribution also 
change the shape of the AO filter, as does imperfect DM calibration.
As a result, the observed AO control radius is often smaller 
(in angle) than the theoretically-predicted AO control radius.

Following the methods in SKMBK, we use a high-pass filter with a power 
law distribution in spatial frequency space to mimic the action of the
AEOS AO system,
	\begin{equation} \label{powerLaw}
	F(k) = \cases{(k / k_{AO})^n, &for\  $k < k_{AO}$\cr
	               1, &otherwise.\cr}
	\end{equation}
In previous work \citep{MakidonSPIE03}, we assumed a `parabolic' filter
(\eg $n=2$) to model this AO system.  However, further refinement
of our simulations have shown us that a $n=0.9$ power law distribution 
best fits the AEOS system (see Section~\ref{sec:AEOSsims}).
We assume single-layer atmospheric turbulence, and generate independent
realizations of Kolmogorov-spectrum atmospheric phase screens in 
$1024\times 1024$ arrays to simulate this turbulence.  We use code
implementing a rapid Markov method for generating  Kolmogorov-spectrum
phase screens \citep{Lane92, Glindemann93, Porro00}, with a spatial scale
of 141.047 pixels per meter in the aperture plane.

We Fourier transform the input phase arrays, and multiply them by the
power law filter in Equation~(\ref{powerLaw}) to mimic the action of AO.
Then we reverse-transform the spatially filtered arrays to obtain the
AO-corrected wave front without waffle error.  We use only the central
half of the smoothed phase screens, in order to avoid any edge effects 
introduced by the finite extent of the phase screens during the 
numerical manipulations.

\subsection{Simulating Waffle Mode Phase Error} \label{sec:SimWaffle}

For traditional Shack-Hartmann AO systems and WFR, waffle mode phase error
present in the atmosphere remains unsensed and uncorrected, and thus leaks
through the system.  This is also true of the spatial filter we use to 
represent our AO correction.  However, because of our code structure, we
found it easier to add the waffle mode phase error back into the AO-corrected
residual phase after correction rather than before.  We found no difference
between these two approaches.

\begin{figure}
	\epsscale{0.95}
	\plotone{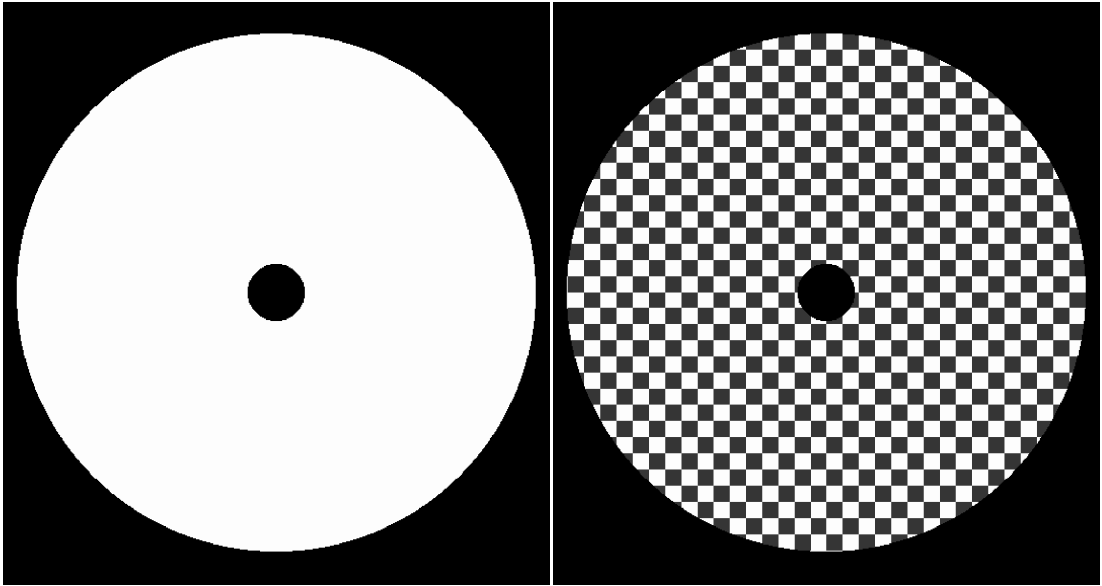}
	\caption{
	The AEOS pupil and subapertures over which we calculate our simulations.
	Left: the telescope input pupil.  The diameter of the telescope is 512
	pixels, with the secondary mirror obstruction subtending 56 pixels.  
	Right: the distribution
	of subapertures over the telescope pupil over which we calculate the mean
	$\phi_{w}$ in the incoming phase screen.
	}
	\label{fig:WafflePupil}
\end{figure}

We calculated the waffle mode error, $\phi_{w}$, present in our incoming
atmospheric phase aberrations $\phi_{atm}$ by first representing the AEOS 
WFS subapertures as a checkerboard pattern over our aperture, with
white and black squares corresponding to the two families of DM actuators 
(see Figure~\ref{fig:WafflePupil}).
We then calculated the mean phase of the incoming wave front
$\phi_{atm}$ over the white squares ($\phi_1$), 
as well as the mean over the black squares ($\phi_2$).
The zero-mean waffle phase error present in the atmospheric phase
aberration is then constructed by  assigning a phase $\phi_{w}$ defined
by $(\phi_1+\phi_2)/2$ over the white squares, and $(\phi_1-\phi_2)/2$ over
the black squares.
We then added $\phi_{w}$ back into our AO-corrected phase arrays 
($\phi_{AO}$), producing an AO-corrected wavefront with waffle mode phase
error.

Since $\phi_{AO}$ is nonzero in this case, the aperture illumination 
function takes the form
\begin{equation} \label{eqn:AOWaffleApertureIllum}
	A_{\psi} = A\,A_{\phi} = 
	    A\,e^{i\left [\phi_{AO}(x,y)+\phi_{w}(x,y)\right ]}.
\end{equation}
The PSF corresponding to this illumination function is just the square of
the absolute value of the Fourier Transform of $a_{\psi}$.

\begin{figure}
	\epsscale{0.95}
	\plotone{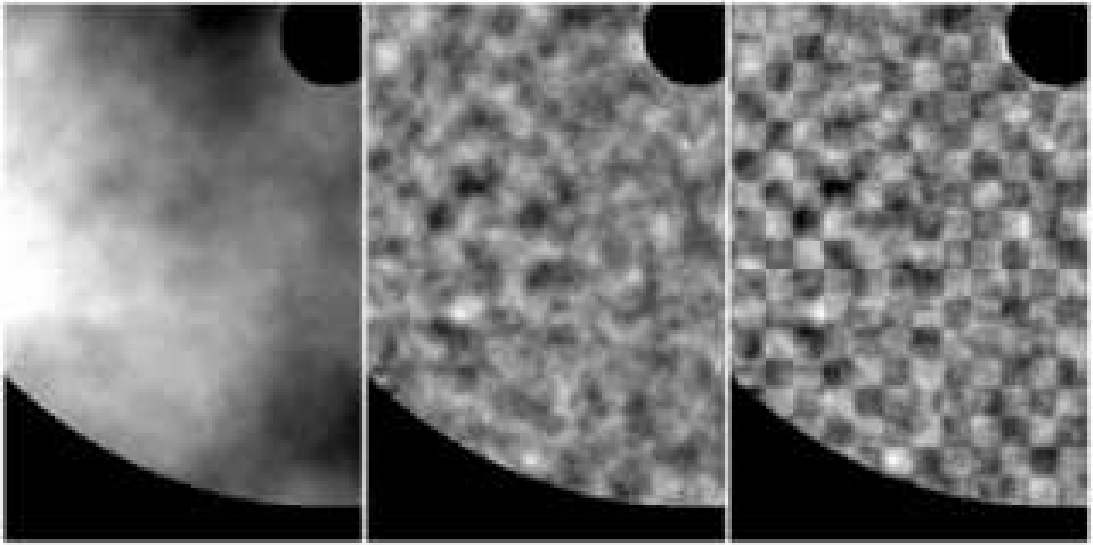}
	\caption{
	Simulated Kolmogorov-spectrum and AO-corrected phase screens over the
	telescope pupil.  Left: the incoming phase screen over a portion of the
	AEOS pupil with linear scaling.  The AEOS secondary obstruction is shown
	at the upper-right.  Center: the residual AO-corrected phase screen over
	the same portion of the AEOS pupil.  Here, the linear scaling has been
	enhanced relative to the figure at left to show the residual structure. 		
	Right: the same residual AO-corrected phase screen as at center
	with waffle mode introduced, shown at the same linear scaling as at center.
	Here, $\phi_{w}=0.03$ radians with ${\cal W}=10$.
	}
	\label{fig:WafflePhase}
\end{figure}

We show this process graphically in Figure~\ref{fig:WafflePhase}, over a 
portion of the AEOS pupil.  We calculate the amount of zero-mean waffle 
phase error present in the incoming phase screen (left) over the checkerboard
array of subapertures.  Without waffle mode phase error, the AO-corrected
wave front would just be dominated by residuals with spatial frequencies
beyond those correctable by our spatial filter.  However, the addition of
waffle mode phase error produces a noticeable checkerboard pattern
in the final, AO-corrected wave front (right).

\subsubsection{Waffle strength a function of $r_0$} \label{sec:PhiVariations}

As part of our analysis, we examined how $\phi_{w}$ varies with the Fried
length $r_0$.  We chose eight values of $r_0$, giving us $D/r_0$  between
5 and 40 for the AEOS geometry, and created 100 independent realizations of
Kolmogorov-spectrum phase screens for each.  We then calculated $\phi_{w}$ 
over the AEOS aperture for each phase screen realization.

\begin{figure}
 \epsscale{0.7}
 \plotone{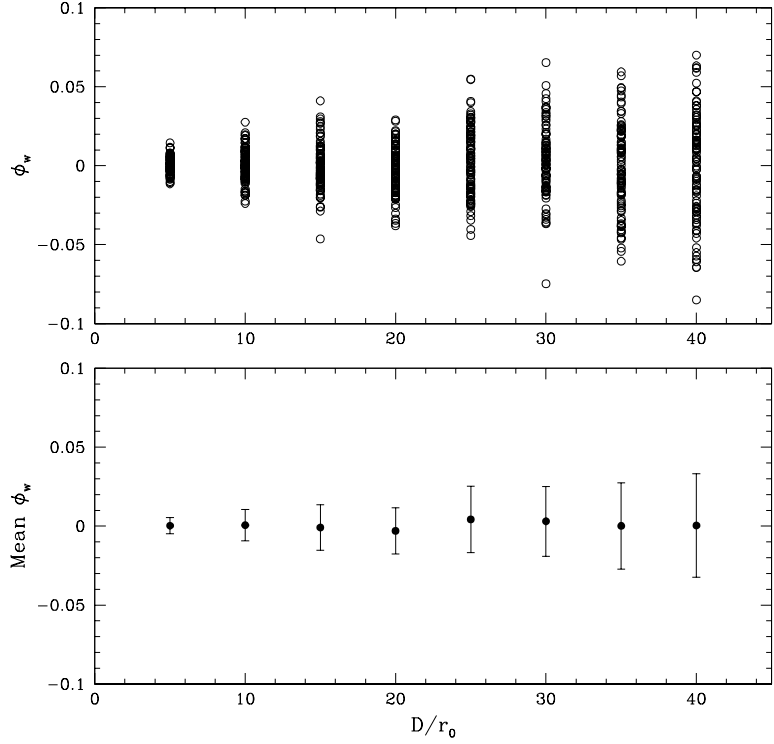}
 \caption{
   Measured mean atmospheric $\phi_w$ values from 100 Kolmogorov-spectrum
   phase screens with the AEOS telescope and wavefront sensor subaperture
   geometries (top), with the mean measured value and standard deviation
   of the mean (bottom).
 }
 \label{fig:WafflePhi}
\end{figure}

Figure~\ref{fig:WafflePhi} shows the expected trend of increasing 
variance of $\phi_{w}$ with increasing $D/r_0$.  For cases where 
$D/r_0$ is greater than the number of WFS subapertures, the mean 
measured $\phi_{w}$ can be as high as $\phi_{w} =\pm 0.1$ radians. 
More typical values are between $\pm 0.04$ radians.  As we will shown 
in section \ref{sec:WaffleStrehl}, waffle mode error starts to cause 
a significant reduction in Strehl ratio when phase variations due to 
waffle grow beyond 0.1 radian.  This suggests that WFRs which do not 
account for changing sky conditions throughout the night or which 
do not penalize waffle modes may not be able to accommodate these variations,
and may allow waffle and waffle-like modes to build up, effectively
enhancing the atmospheric waffle present in the observations.

\subsubsection{Effects of Waffle Mode on Strehl Ratio} \label{sec:WaffleStrehl}

To analyze the intrinsic effects of waffle mode phase error on the 
observed PSF, we assumed perfect AO correction (\ie $\phi_{AO} = 0$) with
an introduced waffle error of $\phi_{w} = 0.01$ radians of
phase difference. This corresponds to a typical value of $\phi_{w}$
present in our simulated Kolmogorov-spectrum phase screens with the 
AEOS WFS subaperture geometry.
The result is a nearly perfect PSF attenuated by the regular pattern
of satellite PSFs characteristic of fundamental waffle mode error.  Here
we note other waffle and waffle-like modes will produce additional 
patterns of satellite PSFs in the image plane (M. Britton 2004, {\it private
communication}). We note in passing that the strengths of these additional
modes will depend on the details of the WFRs used.

\begin{figure}
 \epsscale{0.75}
 \plotone{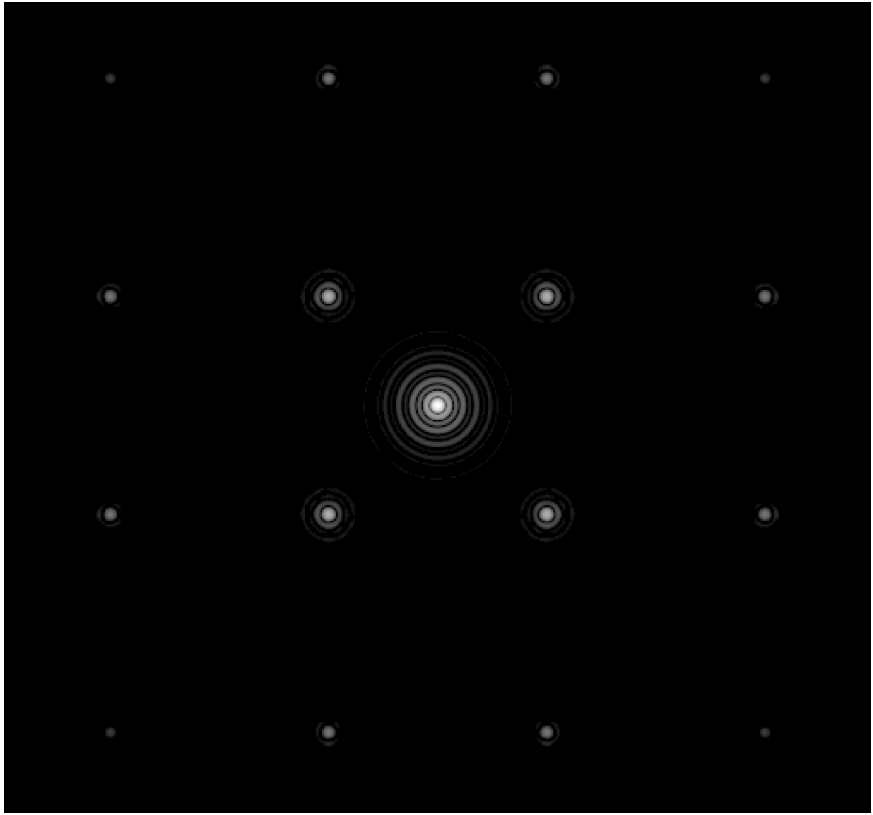}
 \caption{
   Simulated monochromatic image assuming the AEOS telescope and WFS
   geometries with fundamental waffle mode error defined by $\phi_{w}=0.01$ 
   radians and ${\cal W}=50$.
   Atmospheric correction is assumed to be perfect (\ie no residual 
   atmospheric phase error other than waffle).
   This image is on a min-max scaling with a logarithmic stretch.  In this
   extreme case, one can clearly see the primary, secondary, and tertiary
   waffle spots around the central PSF (see text).
 }
 \label{fig:NoAtmosphereWaffle}
\epsscale{1}
\end{figure}

An example of our ``waffle only'' simulations can be seen in 
Figure~\ref{fig:NoAtmosphereWaffle}, where the waffle mode phase
error has been increased to 0.5 radians to show the grid of satellite
waffle PSFs.  We note the primary, secondary, and tertiary waffle 
PSFs appearing at the intersection of grid lines a distance 
$n\lambda / 2d$ (where $n = 1,3,5,\dots$). from the optical axis.  As
discussed in Section~\ref{sec:wafflePSF}, the waffle PSFs which fall the
grid points where $n = 2,4,6,\dots$ hit the zeros of the sampled function
${\rm sinc} (d\theta)$, and do not appear in the final image.

We define a waffle amplitude factor ${\cal W}$ to describe the
enhancement of the waffle mode beyond the atmospheric $\phi_{w}$
calculated from our simulated phase screens.  In reality, ${\cal W}$ 
describes the amplification of the waffle mode due to the choice of 
wave front reconstructor, imperfect WFS flat fields, the
effects of immobile DM actuators, and so on.

We produced a series of PSFs with total phase errors ranging from 
0.01 radian and 1.0 radian (peak-to-valley) by varying ${\cal W}$ between
1 and 100, keeping $\phi_{AO} = 0$ and $\phi_{w} = 0.01$.  We then 
calculated the Strehl ratios of these PSFs by measuring the
peak pixel value of the centroided PSF in the reconstructed aberrated
wave fronts as compared to that from an image formed by a perfect wave
front with the same image plane sampling.

We present our Strehl ratio calculations in Table~\ref{table1}.  We
see little degradation in Strehl ratio for the cases with total waffle
mode phase errors less than $\sim0.10$ radian - Strehl ratios in each
of these examples remained higher than 0.99.  Even for cases with 0.30
radians of total waffle, the calculated Strehl ratio remained above
0.90.  It is only when the waffle mode phase difference grows beyond
this value that pure waffle mode error becomes a major factor in reducing
the Strehl ratio. This suggests that in the absence of other phase errors,
fundamental waffle mode error may not be an important factor in limiting the
correction (and hence the observed Strehl ratio) in AO imaging.  However,
as more high-order AO (and ExAO) systems become prevalent, the use of
WFRs that penalize or limit the growth of waffle modes in the AO system
may be key to maintaining stable PSFs during AO observations.  Additionally,
the development of alternative methods of wave-front sensing, such as the
spatially filtered wave-front sensor \citep{Poyneer04} or the pyramid sensor 
\citep{Ragazzoni96, Verinaud05}, may be required to achieve the contrast
levels necessary to achieve ExAO science.

\subsubsection{Symmetric and Antisymmetric Waffle PSF Terms} \label{sec:WaffleSym}

\begin{figure}
 \epsscale{0.95}
 \plotone{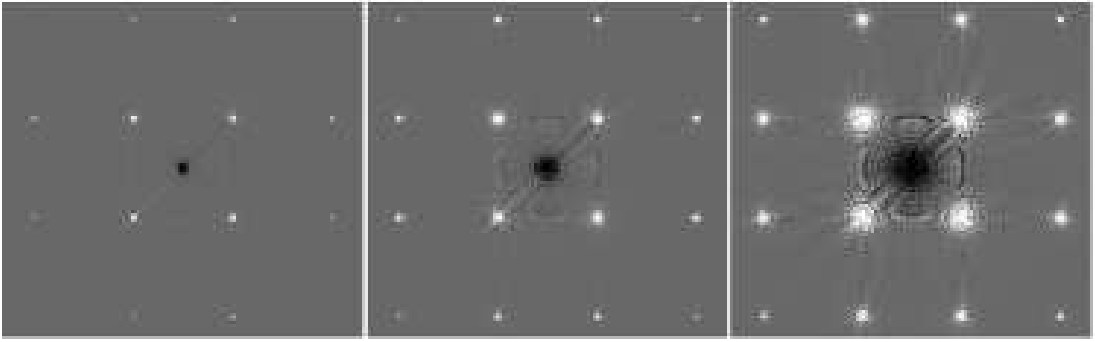}
 \caption{
   Difference between simulated AEOS monochromatic PSFs with and without 
   waffle mode error.  From left to right: PSFs with waffle 
   $\phi_{w} = 0.01,~0.03,~{\rm and}~0.10$ can be seen in the PSFs from
   left to right with calculated Strehl ratios of  99.99\%, 99.91\%, 
   and 99.0\% respectively.
   In each case, atmospheric correction is assumed to be perfect 
   (\ie no residual atmospheric phase error other than waffle).  Subtracting
   the perfect PSF results in a ``hole'' where the on-axis PSF with 
   waffle would be, 
   indicative of a reduced Strehl ratio in the waffle images.  The
   antisymmetric component of the PSF due to waffle mode error is visible,
   though the symmetric component of the waffle mode error dominates.
 }
 \label{fig:WaffleDifferences}
\end{figure}

In \ref{sec:WafflePSFexpansion}, we presented the first two terms of the
PSF expansion of \citet{Sivaramakrishnan02}, and
noted that the symmetric terms of this expansion would dominate the PSF
as waffle mode phase errors increased. To examine this, we made use of
the simulated PSFs described in \ref{sec:WaffleStrehl}, and differenced
these PSFs with a perfect PSF ($p_{0}$), generated with no phase errors.
The result for three of these cases ($\phi_{w} = 0.01,~0.03,~{\rm and}
~0.10$) can be seen in Figure~\ref{fig:WaffleDifferences} for 
aberrated PSFs with calculated Strehl ratios of 99.99\%, 99.91\%, 
and 99.0\% respectively.

We can see that power in the satellite waffle PSFs is accounted for by
the $p_{2, Strehl}$ term,
\begin{equation} \label{eqn:p2_Strehl}
	p_{2, Strehl}(\Phi) = 
	    -\onehalf [a(a^{\ast}~\conv\ \Phi^{\ast}~\conv\ \Phi^{\ast}) +
		           a^{\ast}(a~\conv\ \Phi~\conv\ \Phi)]
\end{equation}
as illustrated by the on-axis ``hole''.  There
is some evidence of the antisymmetric $p_{1}$ term in the diagonal patterns
in these difference images, but their effect on the waffle PSF is small at 
these Strehl ratios.  This antisymmetry is more obvious as Strehl ratios
decrease to between 97.0\% and 90.0\%, but the $p_{1}$ term never 
dominates the intensity distribution.

\subsection{Simulating AEOS Adaptive Optics} \label{sec:AEOSsims}

The AEOS WFS is a Shack-Hartmann system with an array
of $84~\micron $ lenslets, with each lenslet producing a Hartmann spot 
that is imaged onto a $4 \times 4$ group of pixels on the WFS CCD.  The 
deformable mirror is a 31.5 cm diameter Xinetics DM supported by 941 
lead magnesium niobate (PNM) actuators spaced 9 mm apart on a square grid,
with 35 actuators spanning the diameter of the DM \citep{Roberts02}.

Because of initial concerns about pupil wander during AO imaging, early
AEOS wave front reconstructors did not use the entire pupil for 
wave front reconstruction.  Subapertures near the edge of the pupil were
assumed not to be fully illuminated during each observation, and were
considered unreliable
for active wave front reconstruction.  To account for this, we modeled 
33 active actuators ($N_{act} = 33$) across the pupil diameter instead 
of the full 35 actuators across the DM, giving us an undersized DM
with a 28.8 cm clear area and providing us with a projected actuator 
spacing of order 0.11 m/actuator at the primary mirror.  

To mimic finite DM actuator throw, we assumed an actuator stroke limit of
$\pm\,2.00~\micron$.  While our simulations model 855 active actuators on
the primary ($33^{2}~{\rm actuators}\times\pi/4$), the actual AEOS WFRs actively control 810
actuators, with actuators projected near the edges of the primary and
secondary mirrors slaved to neighboring actuators.  We did not model slaved
actuators, nor did we model the effects of actuator influence functions,
non-linear actuator motions, or actuator hysteresis.  We also do not account
for system response time or calibration errors. 

\subsubsection{Seeing Measurements at Haleakala} \label{sec:seeing}

The AEOS Adaptive Optics imaging data we used for our comparison comprises
four sets of twenty-five closed-loop \Iband\ observations of the K3 II star
HR 7525 ($\gamma\,$ Aql, $V\eq 2.72$) taken at two separate exposure times 
(48 ms and 73 ms) with the Visible Imager (VisIm) camera using a 2 magnitude
neutral density filter.  This star also served as the WFS target during 
these observations.  These data were 
acquired on 8 October 2002 and span a total of 5.43 minutes from the
first to the last exposure in the sequence.  The images were processed
(dark, bias, and flat field corrected) using the task CCDPROC in 
IRAF.\footnote{IRAF is distributed by the National Optical Astronomy
Observatories, which are operated by the Association of Universities for
Research in Astronomy, Inc., under cooperative agreement with the National
Science Foundation.}

Measurements of $r_{0}$ taken with the University of Hawaii's Day Night
Seeing Monitor (DNSM) during the night of our observations showed a 
mean measured $r_{0} = 11.44 \pm 3.54~{\rm cm}$ at $\lambda = 0.500\micron$ 
from 320 data points over 9.22 hours.  The median measured $r_{0}$ for this
night was $11.97~{\rm cm}$.  The mean zenith-corrected $r_{0}$ for this
night was $r_{0} = 13.17 \pm 3.88~{\rm cm}$, with a median value of
$13.73~{\rm cm}$.  The nine measurements acquired just before,
during, and after our observations showed a mean zenith-corrected $r_{0}$
of $11.63~{\rm cm}$ with minimum and a maximum measured values of 
$8.84~{\rm cm}$ and $13.60~{\rm cm}$ respectively.

This variation is apparent in our data, as measurements of the full-width
at half maximum (FWHM) of the PSF in each image showed a median of 
$3.17 \pm 0.26~{\rm pixels}$ for 92 images, with a minimum FWHM of 
2.77 pixels and a maximum FWHM of 4.49 pixels.  Over its nominal optical
bandpass, (from $0.70~\micron$ to $1.050~\micron$) the VisIm has a pixel 
scale of $0.0211 \pm 0.004~{\rm arcsec/pixel}$, giving us a median AO 
FWHM of $0.0669~{\rm arcsec}$.  

Unless otherwise noted, we have adopted the value $r_{0} = 11.60~{\rm cm}$ 
to define our nominal seeing conditions. We scale the value of $r_{0}$ \Iband\ 
and to \Hband\ following the relationship
\begin{equation} \label{eqn:r0wavelength}
	r_{i} = r_{0}{\lambda_{i} \overwithdelims () \lambda_{0}}^{6/5} 
	    {\cos^{3/5} {\zeta}}
\end{equation}
where we assume the zenith angle, $\zeta = 0$ \citep{Hardy98}.  In this way,
we calculate $r_{\rm I}\eq 22.92~{\rm cm}$ and 
$r_{\rm H}\eq 48.20~{\rm cm}$, and use these values in our simulations. 

\subsubsection{Estimating Waffle in AEOS \Iband\ Data} \label{sec:AEOSwaffle}

To estimate the amount of waffle present in the AEOS \Iband\ data, we
first required a knowledge of the Strehl ratios of the images themselves.
We measured Strehl ratios by first simulating the diffraction pattern of 
the AEOS telescope, as computed from the analytical
expression for the diffraction pattern of a centrally obscured circular
aperture.  We extracted photometry of the PSF and diffraction pattern,
and registered the PSF and the diffraction pattern to sub-pixel resolution
by cross correlation.  The PSF was then shifted to have the same sub-pixel
center as the diffraction pattern, and the Strehl ratio is computed from
the ratio of peak pixel intensities in both images (see \citet{RobertsSPIE04}
for a discussion of Strehl computation algorithms).  Strehl ratios measured
in this manner ranged from 0.17 to 0.28.

We measured the intensity of the peak pixel in each of the four satellite
``waffle PSFs'' observed in the images of HR 7525, and compared these
measurements to the peak pixel of the on-axis PSF.  We found that the peak
intensity of each waffle PSF was of order 0.25\% the peak intensity of
the on-axis PSF.  The total flux measured in the waffle PSFs was 
$\sim1.6\%$ the flux of the on-axis PSF.

We then compared this result with the relative intensities of the
waffle spots produced through simulation.  Our simulations made use
of five independent realizations of Kolmogorov-spectrum phase screens,
each with $r_0\eq 22.92~{\rm cm}$.  
Our AO correction model used the $0.9$ power law filter in spatial 
frequency space, as described earlier in Section~\ref{sec:simulations}.
The residual phase that remained after AO correction (\ie $\phi_{AO}\neq 0$)
was then used to simulate a PSF for each of the five incoming phase screens.
These PSFs where then co-added to produce our final image.

We simulated a set of monochromatic PSFs using this set of five phase
screens, calculating $\phi_{w}$ from the phase screens themselves.  We
chose waffle amplitudes ${\cal W}$ ranging from zero, to atmospheric, to
the extreme (\eg ${100\cal W}$), and
ran simulations using the same five input phase screens for each value of 
${\cal W}$.  We then examined the resultant images, and assumed the peak
intensity of the simulated waffle PSFs corresponded to the peak wavelength
of the \Iband\ filter transmission function used in the VisIm observations.
We then chose the value of ${\cal W}$ which produced differences in the peak
waffle spot and central PSF intensities which best matched the observed 
values from the data.

Our simulated images are generated at a pixel scale of $\lambda/4{\rm D}$,
or $0.0125~{\rm arcsec/pixel}$, nearly a factor of two finer sampled than
the true VisIm sampling at this wavelength ($0.0211 \pm 0.004~{\rm arcsec/pixel}$).  
We rebinned the simulated data to match the VisIm pixel scaling using the IDL task
FREBIN \citep{Landsman93}.  We added Gaussian detector read noise 
(12 $e^{-}$ rms)  and dark current (22~$e^{-}$~pixel$^{-1}$~s$^{-1}$) into our 
rebinned  PSFs, and modeled VisIm detector charge diffusion and optical effects 
by convolving our simulated images with a Gaussian profile of $\sigma\,=\,1.15$ 
pixels using the IRAF task GAUSS.  This has the effect of limiting the maximum 
observable Strehl ratio in simulated VisIm data to $S\sim0.70$, which matches 
expectations for the VisIm and its optics as used in October 2002.

\begin{figure}
	\epsscale{0.95}
	\plotone{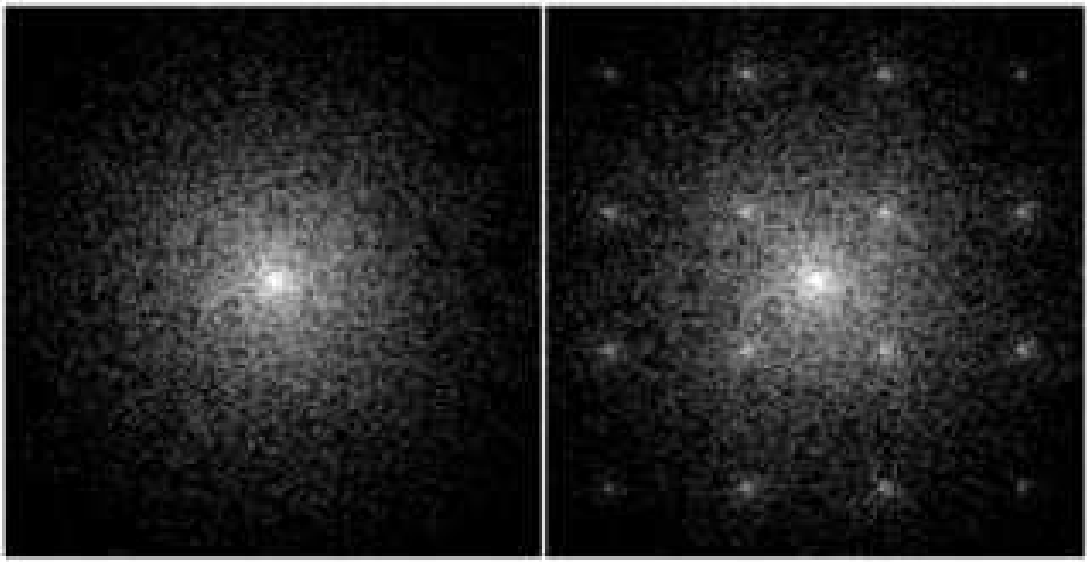}
	\caption{
	  Simulated AEOS monochromatic PSFs without and with waffle mode
	  error, displayed on the same logarithmic scaling. These images assume
	  $D/r_{0}=32$ at $\lambda=0.88~\micron$ with a power law AO correction 
	  with exponent 0.9
	  (see text).  Neither read noise nor Poisson noise are included here.
	}
	\label{fig:WaffleImages}
\end{figure}

We found a waffle mode amplitude of $16{\cal W}$ at  $\lambda\eq 0.88~\micron$
best matched the intensity of waffle spots in these VisIm \Iband\ images.
An example of this monochromatic simulation with full
AO correction, both without and with waffle, can be seen in
Figure~\ref{fig:WaffleImages}. We used this as the starting point to extend
our monochromatic simulations to full \Iband\ simulations.

\subsubsection{Simulating \Iband\ Images} \label{sec:AOsim}

We simulate a set of \Iband\ images by first generating thirty-six
monochromatic images across the bandpass, from $\lambda\eq 0.700~\micron$
to $1.050~\micron$ (the effective cutoff wavelength of the VisIm detector) 
in increments of $0.01~\micron$.  We use the same
Kolmogorov-spectrum phase screens for each image at a given $\lambda_{i}$,
scaling the phase difference of each phase screen relative to the
bandpass central wavelength \eg $\phi_{atm}\cdot (\lambda_{i}/\lambda_{ref})$,
where $\lambda_{ref}$ is the peak wavelength of the filter transmission
function in the bandpass. 
We maintained the same 0.9 power law AO filter in spatial frequency space as
before, with $N_{act}\eq 33$ actuators across the primary and a waffle mode 
amplitude of $16{\cal W}$. We did not consider the effects of slaved or
dead actuators as part of our simulations, but did include a 
$\pm 2.0~\micron$ DM stroke limitation in our analysis.

We rebinned the pixel scale of each monochromatic PSF by the ratio of the
simulated wavelength to the reference wavelength to account for the
wavelength-dependent resolution changes inherent in Fourier methods.  
We then multiplied each rebinned $\lambda_{i}$ PSF by the transmission
expected for
a standard Johnson/Bessell \Iband\ filter transmission function obtained
from SYNPHOT in STSDAS,\footnote{SYNPHOT and STSDAS are products of the Space
Telescope Science Institute, which is operated by AURA under contract to
the National Aeronautics and Space Administration.}
and combined the set of monochromatic PSFs for a given phase screen into a
single broadband PSF using the IRAF task IMCOMBINE.

We generated five broadband PSFs in this manner, and
co-added these PSFs into a single ``short exposure PSF.''  Assuming 
a speckle correlation time of order a few tens of milliseconds, these five
PSFs, when averaged, would approximate a $\sim 100 {\rm msec}$ PSF.  We note
that longer exposures can be simulated in this way by adding statistically
independent realizations of PSFs together, but would lack any correlated or
longer-term effects (such as quasi-static speckles) without the addition of
static sources of phase errors.

\begin{figure}
	\epsscale{0.95}
	\plotone{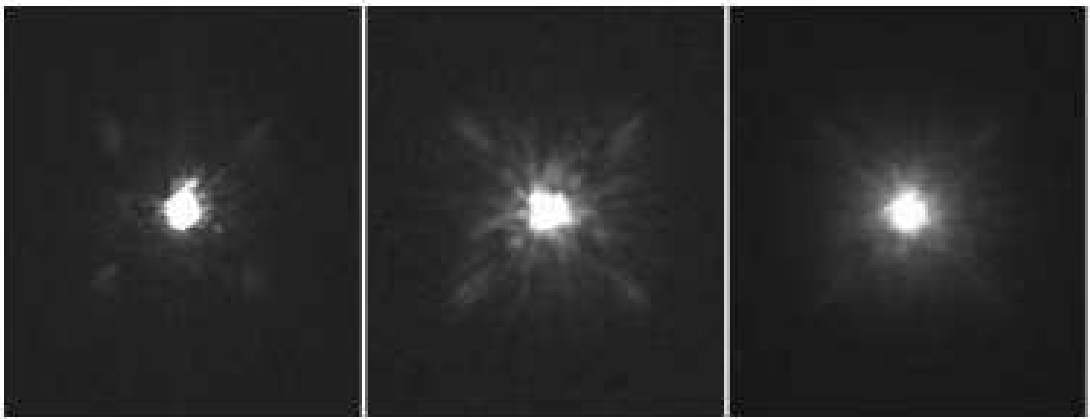}
	\caption{
	At left, an AEOS Adaptive Optics \Iband\ image of HR7525 acquired with
	the Visible Imager (VisIm) on 08~October~2002~(UT).  Strehl ratio
	of this image has been estimated to be $\sim20\%$.  At right, a
	simulated AEOS Adaptive Optics \Iband\ image assuming waffle mode
	error.  The Strehl ratio of the simulated image was calculated to be
	$\sim60\%$ before convolution with a Gaussian filter and and the
	addition of Poisson noise, which reduced the calculated Strehl to
	$\sim42\%$.  Other features in the observed PSF can be explained by
	the addition of phase errors in the input pupil.
	}
	\label{fig:IbandDataSims}
\end{figure}

With the methods described above,
our our simulated Kolmogorov wavefronts produce PSFs with Strehl ratios
between $0.544$ to $0.635$ with purely atmospheric phase disturbances.
When waffle mode phase error with ${\cal W}\eq 16$ is introduced, the
Strehl ratios decrease to between $0.406$ and $0.556$ for the same phase
screens.  When convolved with the Gaussian filter, the Strehl ratios of our
simulated PSFs were further reduced by $70\%$, to between $0.284$ and 
$0.389$.  However, the resultant final images produced waffle spots with the 
same morphology and intensity distribution as in VisIm images (Figure~\ref{fig:IbandDataSims}). 
The total flux in the visible waffle spots, $\sim1.8\%$ the flux of the
central PSF, is in good agreement with the $\sim1.6\%$ value measured in
the VisIm data.  The peak intensities of each waffle spot, $\sim 0.25\%$ of
the peak of the central PSF, is also in agreement with the VisIm data.

While the simulated Strehl ratios remain higher than expected for the 
$r_0$ values observed in the VisIm images from this night, we note that our
simulations only account for two sources of phase error: residual 
atmospheric variations and the fundamental waffle mode.  We do not account 
for pointing variations or pupil wander in the telescope beam, nor do we
employ a realistic model of the DM.  \citet{Abreu00b} noted that
pinned and slaved actuators may provide as much as a $10\%$ hit to the 
observed Strehl, while actuator influence functions and actuator dynamics 
will also limit the quality of the wavefront reconstruction.  Any observed
non-common path aberrations between the AEOS WFS/DM and the VisIm will also
degrade the observed Strehl ratios, as will the effects of the VisIm detector
pixel  modulation transfer function (MTF).

We stress here that these VisIm images, acquired early in the AEOS AO system
lifetime, did not benefit from the  waffle-suppressing  WFRs that are now
in use at AEOS (G. Smith 2003, {\it private communication}).  
Images obtained more recently in $I-$, $J-$, $H-$, and $K-$bands have shown
waffle mode error in AEOS imaging to be at or near nominal (atmospheric) 
levels \citep{PerrinAMOS03}.

\subsubsection{The Magnitude of Waffle Mode Phase Error} \label{sec:waffleRMS}

Following \citet{Sandler94}, we note that for Strehls $\gtrsim 0.30$ the 
final observed Strehl ratio of a system can be considered as a multiplicative
combination of Strehl ratios due to successive uncorrelated sources of error.
 Thus 
\begin{equation} \label{StrehlHit}
	S = S_{\phi_1} S_{\phi_2} \dots S_{\phi_n} =		
		e^{-\sigma^2_{\phi_1}}\,e^{-\sigma^2_{\phi_2}}\,\dots\,
		e^{-\sigma^2_{\phi_n}}
\end{equation}
where $\sigma^2_{\phi_1}$ is due to one class or family of aberrations
(\ie high-order aberrations), while $\sigma^2_{\phi_2}$ is due to
another class of aberrations uncorrelated with the first, etc.

Neglecting the optical effects of the VisIm, we can calculate the mean square
wavefront error for atmospheric phase disturbances, $\sigma^2_{\phi_{AO}}$,
from the mean Strehl ratio of our simulated images,
\begin{equation} \label{StrehlAO}
	S = S_{\phi_{AO}} =	e^{-\sigma^2_{\phi_{AO}}} = 0.589.
\end{equation}
If we then assume $\sigma_{\phi_{AO}}\eq 2\pi z / \lambda$, where
$\lambda\eq 0.880\micron$, we find $z$ -- the residual RMS
wavefront of the AO-corrected atmospheric phase screen -- to be 
$101.9~{\rm nm}$. If we then consider the addition of waffle mode
phase error with ${\cal W}\eq 16$, we find,
\begin{equation} \label{StrehlAOwaffle}
	S = S_{\phi_{AO}} S_{\phi_{w}} = \\
		e^{-\sigma^2_{\phi_{AO}}} e^{-\sigma^2_{\phi_{w}}} = 0.488.
\end{equation}
With $S_{\phi_{AO}}\eq 0.589$ as above, $S_{\phi_{w}}\eq 0.829$, and
thus $\sigma^2_{\phi_{w}}\eq 0.188$.  Following the method above, we
calculate the residual RMS wavefront due to waffle mode phase error
to be $z_{w}\eq 60.7~{\rm nm}$.  
\newpage
\subsection{Other Potential Source of Waffle Mode Phase Errors} \label{sec:failure}

While our simulations were able to reproduce the general morphology of
the early AEOS AO system PSF as well as the shape and energy seen in
the primary waffle PSFs, we realize our work does not fully reproduce
the AEOS optical system, WFS geometry, or WFRs.  In fact, the presence of
immobile DM actuators in the AEOS pupil imprints a phase and an amplitude
error across the entire pupil.  \citet{Abreu00b} has estimated the Strehl
hit to AEOS AO system performance due to these
immobile actuators at $\sim10\%$.  Initial work has suggested these failed
actuators can mimic localized waffle modes described by \citet{GavelSPIE03}.

Given a DM surface map taken with an interferometer, it is relatively
straightforward to calculate the waffle present at the DM surface.  Comparing
this to the strength of the waffle seen in real images will indicate whether
WFRs tailored to minimizing the effects of failed actuators will
help to significantly improve the image quality.  A reconstructor which 
drives the wave front to the best-fitting plane through the failed 
actuators' heights on the DM is one obvious strategy.  However, such a 
strategy would adversely affect the usable stroke of the DM, and thus would
only be beneficial only when the seeing is good.

Circular aperture boundaries and partial subaperture obscuration by
edges and secondary support spiders will affect the equations relating
WFS slopes to actuator commands. Other waffle modes can be introduced
by secondary support structure obscurations.  In the extreme case, a
completely obscured line of subapertures partitioning the pupil 
into disconnected sections can result in free-floating piston between 
the sections in the reconstructed wave front. While this form of 
error might be less obvious, it can adversely affect Strehl ratios 
and dynamic range of companion searches and debris disk studies 
because of its proximity to the core of the reconstructed PSF. These 
quadrants must somehow be stitched together explicitly by the 
reconstructor algebra with the mathematical equivalent of extra 
equations of constraint. Partially-illuminated subaperture data 
should be ignored or used differently from WFS centroid data from 
fully-illuminated apertures in order to reduce their contribution 
to undesirable behavior of the reconstructor.

\section{Predicting AEOS AO System Performance for Near-Infrared Observing} \label{sec:NIR}

In order to understand the potential capabilities of diffraction-limited
near-infrared coronagraphy on AEOS, we extrapolated our \Iband\ 
simulations to \Hband\, employing the same power-law filter in spatial
frequency space to simulate wave front correction as we did for our \Iband\
simulations.  The use of the same AO representation provides a reasonably
accurate simulation of a facility AO system working independently of 
the science imaging instrument, as in the case of the AEOS AO system
(among many others).  Here, we discuss details of our \Hband\ simulations. 

We generated realizations of Kolmogorov-spectrum phase screens 
with  $r_{0}\eq 48.20~\rm{cm}$, which created a set of \Hband\ phase screens
with an effective $D/r_0$ of 7.5. We assumed a waffle amplitude of 
$16{\cal W}$ as before, with a DM stroke limit of $\pm\,2.00~\micron$, and 
ran a set of monochromatic simulations covering the wavelength range
from $1.43~\micron$ to $1.83~\micron$ in increments of $0.01~\micron$.
We combined this set of monochromatic images into a single broadband image
(with the appropriate rebinning, described in \ref{sec:AOsim}),
weighting each wavelength's PSF with the Mauna Kea Observatories
(MKO) \Hband\ filter transmission function \citep{Simons02}.
We did not include any other optical characteristics of the telescope,
AO system, or imaging system for these simulations.

\begin{figure}
	\epsscale{0.95}
	\plotone{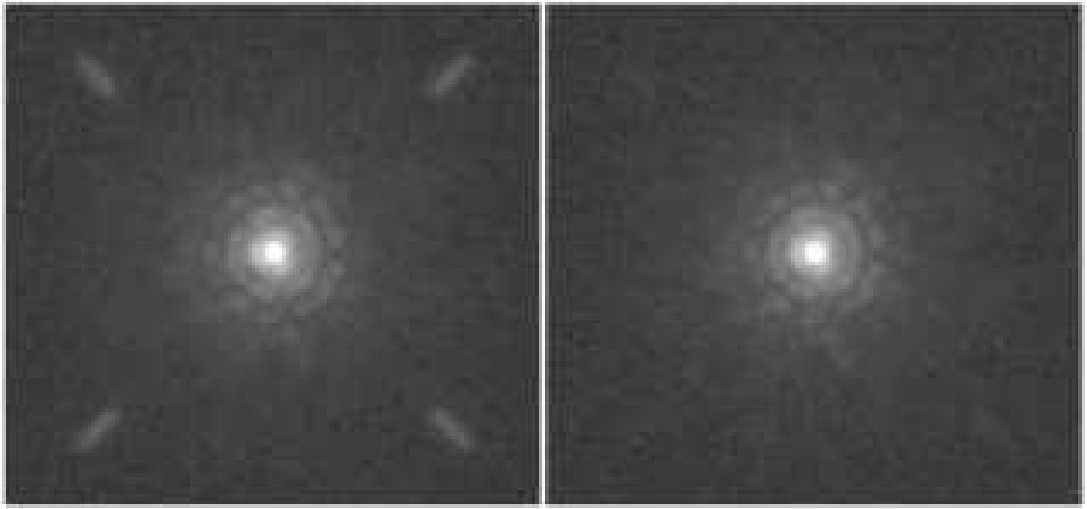}
	\caption{Simulated AEOS \Hband\ PSFs.  Both of the PSFs shown here were
	co-added from five individual PSFs, each using a separate realization 
	of Kolmogorov-spectrum phase screens with $r_0\eq\ 45.6{\rm\ cm}$ as input.
	At left is a PSF with ${\cal W}\eq\ 16$, with ${\cal W}\eq\ 4$ for the
	PSF at right.  Poisson noise and detector read noise have been added to
	these PSFs.  Calculated Strehl ratios are 0.79 for the PSF at left,
	and 0.82 for the PSF at right.}
	\label{fig:Hband2Waffles}
\end{figure}

Our initial simulations suggested the AEOS AO system is capable of high
Strehl ratio images in the \Hband\ even without improved WFRs 
\citep{MakidonSPIE03}, though the presence of significant
satellite waffle spots could make faint companion searches problematic in
particular regions relative to a bright AO target star (see
Figure~\ref{fig:Hband2Waffles}).
However, recent improvements to the AEOS WFRs (G. Smith 2003, 
{\it private communication}) have reduced the waffle mode error remaining
in AEOS imaging to levels that are almost undetectable in typical images.

In March 2003, $J-,$ $H-,$ and \Kband\ images of several stars were taken
with the Kermit Infrared Camera as part of a study of atmospheric turbulence.
This data presented us an opportunity to validate our simulations against
actual data.  The camera and data set up are detailed in \citet{PerrinAMOS03} . Our
simulations with a waffle amplitude of $16{\cal W}$ proved to be too aggressive
for the AEOS WFRs in use at the time of the Kermit observations, and we have
since re-determined our ``best'' ${\cal W}$ for AEOS \Hband\ imaging to be
closer to $4{\cal W}$, well into the noise of most imaging systems.

\begin{figure}
	\epsscale{0.95}
	\plotone{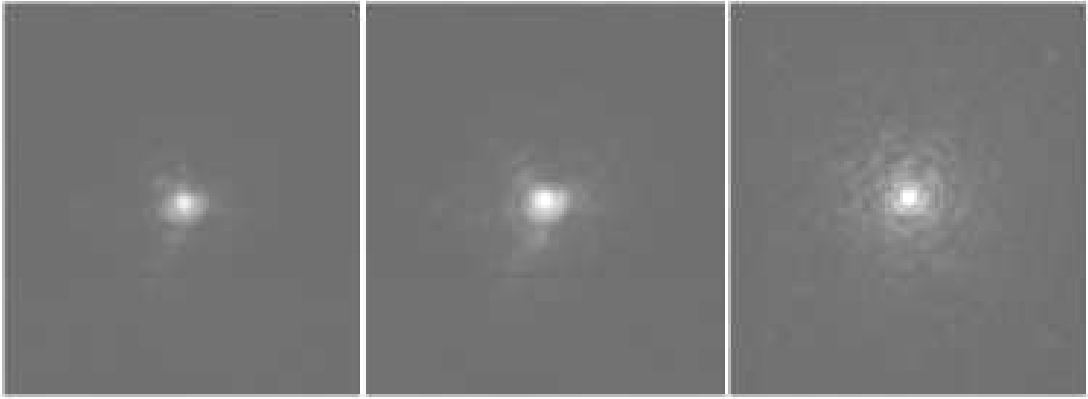}
	\caption{
	Kermit Infrared Camera short-\Hband\ images of the G8III star HR~4667
	(HD~106714, HIP~59847)
	acquired on 16 April 2003 with AEOS Adaptive Optics.  At left, a 
	single 1.5 second exposure with a measured Strehl of $\sim 0.51$.  A 5.0
	second exposure of the same star is shown at middle.  At right is a
	simulated PSF with a single-layer Kolmogorov-spectrum phase screen
	with $D/r_0 \eq 15.0$, with a calculated Strehl ratio of $\sim 0.55$.
	Since our simulations assume perfect optical surfaces, the calculated
	Strehl ratios for the simulated PSF are higher than those imaged with
	Kermit on the night of the observations.
	}
	\label{fig:KermitHband}
\end{figure}

On-the-sky results using the Kermit Infrared Camera have since demonstrated 
the improvements to the AO system, showing only the faintest traces of waffle 
mode phase error in over-exposed images.  
In Figure~\ref{fig:KermitHband}, we show Kermit short-\Hband\ data of the
G8III star HR~4667 (V\eq 4.94) acquired at AEOS on 16 April 2003.  Images
shown were taken with two exposure times: 1.5 seconds (at left), and 5.0
seconds (middle), with measured Strehl ratios of order $\sim0.50$.  Here, we
found that Kolmogorov-spectrum phase screens 
with  $r_{0}\eq 24.2~\rm{cm}$ at \Hband\ best matched the general morphology
and Strehl ratios of these images, suggesting the seeing on the night these
data were acquired was below nominal.  However, we note that these 
relatively long exposures will undoubtedly contain many tens of speckle
lifetimes, while our simulated images only use five independent realizations
of a simple atmosphere.  As such, atmospheric speckles will tend to be
``washed out'' in the real data, but instrumental effects will still be visible.

\citet{PerrinAMOS03} reported measured \Hband\ Strehl ratios as high 
as 83\% during good seeing conditions at AEOS.  Our simulations match these
Strehl ratios in the absence of
other aberrations, though are yet unable to reproduce the quasi-static
speckle structure present in the AEOS \Iband\ and \Hband\ images.  In the
future we hope to be able to determine phase and amplitude maps for the AEOS
AO system beam as it is presented to the Lyot Project coronagraph and to the 
Kermit Infrared Camera to enable proper PSF calibration both with and
without the coronagraphic occulting and Lyot stops in place.
\newpage
\section{Conclusion} \label{sec:conclusion}

One of the goals of this work is to show the utility of simulating
AO system performance in a simple manner, rather than employing
detailed, component-by-component modeling, to enable long exposures
to be simulated in a short time using modest computing power.
We presented the AEOS telescope and its AO system
as a specific case of interest, but note this method should work on
almost any system given imaging data with which to calibrate 
the spatial filter model of AO. Other software tools, such as PAOLA 
(Performance of Adaptive Optics for Large Apertures, \citet{Jolissant04}), 
incorporate more physics into such ``fast'' methods of AO system 
simulation.  These types of simulations are useful both for highlighting
areas where the AO system performance could be
improved, and for predicting instrument performance to 
investigate observing scenarios to investigate what science is
within the scope of an AO system and its science camera.

AEOS and its AO system can prove to be an excellent testbed for
technologies requiring diffraction-limited observing. A better
understanding of the AO system, such as that attempted by this
analysis, will allow the astronomical community to build instruments to
take better advantage of the unique capabilities of AEOS. Studies like
these will provide the astronomical community an opportunity to learn
lessons which can be applied to high-order AO systems on 8 and 10 m
class telescopes. 

\acknowledgments 
 
The authors are grateful to D.~T.\ Gavel, G.~Smith, and P.~E.\ Hodge
for helpful discussions, and to the the staff of the Maui Space
Surveillance System for their assistance in taking these data.  We are
also grateful to to L.~W.\ Bradford, T.~A.\ ten Brummelaar, J.~R.\ Kuhn, 
N.~H.\ Turner, and K.~Whitman for the use of the Kermit image used 
in this paper. 
The authors wish to thank the anonymous referee for suggestions which
have improved this manuscript.
The authors also wish to thank the Space Telescope Science Institute's
Research Programs Office, its Director's Discretionary Research Fund,
and its generous Visitors' Program.
R.S. and M.D.P. are supported by NASA Michelson Postdoctoral and Graduate
Fellowships, respectively, under contract to the Jet Propulsion Laboratory
(JPL) funded by NASA.  The JPL is managed for NASA by the California
Institute of Technology.
The research presented here was supported by the STScI Director's 
Discretionary Research Fund, NSF grants AST-0088316 and AST-0215793, 
and AFRL/DE through Contract Number F29601-00-D-0204.  Our work has also 
been supported by  the National Science Foundation Science and Technology 
Center for Adaptive Optics, managed by the University of California at 
Santa Cruz under cooperative agreement No.  AST - 9876783.
This research made use of the SIMBAD database, operated at CDS, 
Strasbourg, France.

\clearpage

\begin{deluxetable}{ccccccc}
\tablewidth{0pt}
\tablecaption{Average Strehl Ratio for $\phi_{w}=0.01$ Radians \label{table1}}
\tablehead{
\colhead{Waffle ($\cal W$)} & \colhead{$\phi_{total}$ (rad)} & \colhead{Strehl\tablenotemark{a} (\%)} & \colhead{ } & \colhead{Waffle ($\cal W$)} & \colhead{$\phi_{total}$ (rad)} & \colhead{Strehl\tablenotemark{a} (\%)}
}
\startdata
\phn0.0     &	0.00	&	1.0000 & \  & 20.0    &   0.20    &0.9605\\
\phn1.0	    &	0.01	&	0.9999 & \  & 25.0    &   0.25    &0.9388\\
\phn3.0		&	0.03	&	0.9991 & \  & 30.0    &   0.30    &0.9127\\
\phn5.0		&	0.05	&	0.9975 & \  & 50.0    &   0.50    &0.7702\\
10.0    &	0.10	&	0.9900 & \  & 75.0    &   0.75    &0.5354\\
15.0    &	0.15	&	0.9776 & \  & \llap{1}00.0   &   1.00    &\phn0.2919 
\enddata
\tablenotetext{a}{Strehl ratios in Table~\ref{table1} were computed by taking the ratio of the peak intensity in the aberrated image with that of a perfect image.}
\end{deluxetable}

\end{document}